\documentclass[aps,prb,twocolumn,groupedaddress]{revtex4}
\usepackage{graphicx}
\usepackage{amsmath} 
\usepackage{amssymb}
\usepackage{dsfont}
\begin{document} 
\title{Self-consistent description of Andreev bound states in Josephson quantum dot devices}

\author{Tobias Meng} 
\affiliation{Institut N\'{e}el, CNRS \& Universit\'{e} Joseph Fourier, 
25 avenue des Martyrs, 38042 Grenoble, France}
\author{Pascal Simon} 
\affiliation{Laboratoire de Physique et Mod\'elisation des Milieux
             Condens\'es, CNRS and Universit\'e Joseph Fourier, BP 166, 
             38042 Grenoble, France,}
\affiliation{Department of Physics, University of Basel, 
             Klingelbergstrasse 82, 4056 Basel, Switzerland}
\affiliation{Laboratoire de Physique des Solides, CNRS UMR-8502 Universit\'e Paris Sud, 91405 Orsay Cedex, France}
\author{Serge Florens} 
\affiliation{Institut N\'{e}el, CNRS \& Universit\'{e} Joseph Fourier, 
25 avenue des Martyrs, 38042 Grenoble, France}

\date{\today}

\begin{abstract}
We develop a general perturbative framework based on a {\it superconducting
atomic limit} for the description of Andreev bound states (ABS) in interacting quantum 
dots connected to superconducting leads. A local effective Hamiltonian for dressed ABS, 
including both the atomic (or molecular) levels and the induced proximity effect on 
the dot is argued to be a natural starting point. A self-consistent expansion in 
single-particle tunneling events is shown to provide accurate results even in regimes 
where the superconducting gap is smaller than the atomic energies, as demonstrated by 
a comparison to recent Numerical Renormalization Group calculations. This simple
formulation may have bearings for interpreting Andreev spectroscopic
experiments in superconducting devices, such as STM measurements on carbon 
nanotubes, or radiative emission in optical quantum dots. 
\end{abstract}

\maketitle

\section{\label{sec:intro}Introduction}








When a quantum dot is connected to superconducting electrodes, the proximity
effect drastically modifies the dot's electronic structure due to the local 
formation of Cooper pairs. The density of states on the dot thus exhibits a gap, 
so that the formation of discrete sub-gap states arises \cite{josephson_current_beenakker-houten}. 
These Andreev bound states (ABS) play certainly an important role as they may 
contribute a large part of the spectral weight \cite{NRG_spectral_Bauer} and 
carry most of the Josephson current.\cite{Current_benjamin} 
A physical understanding of the ABS requires to characterize how these states are 
connected to the atomic (or molecular) levels of the uncoupled quantum dot, and to describe 
quantitatively their evolution as a function of several parameters, such 
as gate voltage, Coulomb interaction, tunnel couplings, and superconducting gap.
Whereas the ABS have been observed in metal-superconductor hybrid
structures \cite{ABS_spectrum_rowell}, no direct spectroscopy has so far been
achieved in quantum dot systems. Andreev bound states come in pairs, one state
above and one below the Fermi level, forming a two level system. Consequently,
recent interest in the spectroscopy of the bound states has also been stimulated by 
proposals to use the latter as a qubit. \cite{abs_qubit} At present, several
routes have been suggested, such as STM measurements on carbon nanotubes, 
\cite{Governale1} microwave cavity coupling, \cite{spectrum_fogelstrom} 
visible light emission using a Josephon diode, \cite{josephson_diode}
or noise experiments. \cite{Sadovskyy}

Experimentally, superconducting quantum dots can be realized with carbon nanotubes junctions
or semiconducting InAs islands.
It has been shown that quantum dots connected to superconducting electrodes can be tuned
from a Coulomb blockade regime, to a Kondo regime,\cite{Kondo_CNT_lindeloff,
Kondo_conduction_sclaing_buizert, Kondo_vs_supra_buitelaar-schonenberger} to a
weakly interacting Fabry-Perot regime by changing local gate voltages.
\cite{Fabry_perot_jorgensen-flensberg} The Josephson current at zero
bias and multiple Andreev reflections at finite bias voltage have been measured
in such devices.\cite{Fabry_perot_jorgensen-flensberg,
nanotubes-supercurrent_kouwenhoven, MAR_buitelaar-schonenberger,
proximity-eff_Bouchiat} The transition from a $0$-junction to a $\pi$-junction,
namely a reversal in the sign of the Josephson current, \cite{pert_tunnel_glazman-matveev} 
has also been been observed when a magnetic moment forms on the dot. 
\cite{0_pi_Klapwijk, QD_0_pi_kouwenhoven,wernsdorfersquid, Preprint_bouchiat_schonenberger} 
As a possible application of superconducting junctions, nano-SQUID devices have
also been fabricated. \cite{wernsdorfersquid} 

An exact theoretical description of a quantum dot coupled to superconducting
leads is only possible when the Coulomb interaction is fully neglected. Hence the
interacting single dot system, as described by the Anderson model with
superconducting electrodes, has been so far analyzed by treating the Coulomb
interaction with various analytical schemes, such as the mean field 
theory, \cite{meanf_arovas, transport_qd_supra_Zaikin,meanf_dellanna} the 
perturbation expansion in the Coulomb interaction \cite{pert_U_vecino-yeyati} or 
in the tunnel coupling. \cite{pert_tunnel_glazman-matveev} Non-perturbative
calculations, using the Non-Crossing Approximation (NCA), \cite{NCA_Clerk, NCA_Ando}
or the functional Renormalization Group (fRG), \cite{NRG_current_Karrasch} as
well as numerical simulations based on the numerical renormalization group (NRG), \cite{NRG_Oguri_Tanaka,
NRG_choi, NRG_spectral_Bauer, NRG_energy_Ohashi, NRG_current_Karrasch, NRG_spectral_Hecht}
or Quantum Monte Carlo \cite{QMC_Egger, QMC_Tanuma} have also been developped.

None of the analytical approaches mentioned above is able to describe entirely the physics 
of a quantum dot coupled to superconducting leads. Whereas lowest order perturbation 
expansions in the tunnel coupling will hardly capture the proximity effect induced by the
electrodes~\cite{Governale2}, mean field and weak-interaction approaches will miss the Kondo effect. 
NRG calculations on the other hand can capture the physics of such a system over a 
wide range of parameters, but are numerically demanding and not easily portable to 
more complex molecular systems. More importantly, in the view of describing the
ABS alone, none of these techniques does provide a simple physical picture. Henceforth
we will develop in this paper a new perturbative approach based on an effective 
local Hamiltonian for dressed ABS, that extends the limit of large
superconducting gap proposed previously \cite{H_eff_original_affleck,
H_eff_original_arovas}, which was used by many authors.
\cite{Delta_inf_assym_hewson,NRG_spectral_Bauer,vecino_yeyati,Zazu1,Zazu2,Governale1,Governale3,josephson_diode} 
This approach will illuminate the nature of the ABS in interacting quantum dots,
which can be generally viewed as renormalized superconducting atomic states.
This will provide as well a simple analytical framework that is accurate in the most relevant
cases, and that may thus be useful for interpreting future spectroscopic experiments. 
In particular, calculations provided in the proposals of Refs. \onlinecite{Governale1} and 
\onlinecite{josephson_diode} only {\it qualitatively} apply in realistic situations where 
the gap is comparable or smaller that the atomic energies, even when the gap is large
compared to the hybridization to the electrodes. This interesting regime is precisely the one 
that we want to address in the present work.
In addition, we note that our formalism, which incorporates the atomic (or
molecular) levels from 
the outset, can easily be extended to describe more complex systems, as for 
instance superconducting double quantum dots or molecules with more complicated orbital 
structure (see e.g. Refs. \cite{Double_dot-Oguri, Double_dot-loss, Double_dot-Lee, 
Double_dot-Bergeret}).

We organize our paper as follows. In Sec. \ref{sec:theoretical_form}, the system
is mapped onto an effective local Hamiltonian, similarly to the widely used
atomic limit, but including the proximity effects due to the superconducting leads. 
In Sec. \ref{sec:pert}, the perturbation theory around this limit is set up and
self-consistent equations for the ABS energy renormalizations are derived in order
to extend the validity of the bare perturbative approach. Sec. \ref{sec:results} 
illustrates how this expansion can describe ABS in superconducting quantum dots 
over a wide range of parameters, by a comparison to available NRG 
data. \cite{NRG_spectral_Bauer, NRG_energy_Ohashi}

\section{\label{sec:theoretical_form}Theoretical formulation}
\subsection{\label{subsec:model}Model}

We focus in this paper on a single-level quantum dot coupled to superconducting leads,
which is relevant experimentally for molecular junctions with large single-electron 
level spacing.
A simple Hamiltonian able to describe such a system is given by the
superconducting Anderson model

\begin{equation}
\label{eq:Hamiltonian_complete_normal}
H = \sum_{i = L,R}{H_i^{}} + H_d + \sum_{i = L,R}{H_{T_i}} \text{ ,}
\end{equation}
where

\begin{eqnarray}
		\label{eq:H_lead} 
		H_i^{} &=& \sum_{\vec{k},\sigma}{\epsilon_{\vec{k}}^{} \, c_{\vec{k},\sigma,i}^{\dagger} c_{\vec{k},\sigma,i}^{}} - \sum_{\vec{k}}{\left( \Delta_{i}^{} \, c_{\vec{k},\uparrow,i}^{\dagger} c_{-\vec{k},\downarrow,i}^{\dagger} + \mathrm{h.c.}\right) } \nonumber \\
		\label{eq:H_dot}
		H_d^{} &=& \sum_\sigma{\epsilon_d^{} \, d_{\sigma}^{\dagger} d_{\sigma}^{}} + U n_{\uparrow}n_{\downarrow} \nonumber \\
		\label{eq:H_T_LD}
		H_{T_i}^{} &=& \sum_{\vec{k},\sigma}{\left( t_{}^{} \, d_{\sigma}^{\dagger} c_{\vec{k},\sigma,i}^{} + \mathrm{h.c.}\right) } \nonumber \text{ .}
\end{eqnarray}
In the above equations, $d_{\sigma}^{}$ is the annihilation operator of an
electron with spin $\sigma$ on the dot, $c_{\vec{k},\sigma,i}^{}$ that of an
electron with spin $\sigma$ and wave vector $\vec{k}$ in the lead $i= L,R$, and
$n_{\sigma} = d_{\sigma}^{\dagger}d_{\sigma}^{}$. The leads are assumed to be
described by standard s-wave BCS Hamiltonians $H_i$ with superconducting gaps
$\Delta_{i}^{} = \Delta \, e^{i\varphi_i}$. The phase difference of the latter
is noted $\varphi = \varphi_L - \varphi_R$. Furthermore, the leads are assumed
to have flat and symmetric conduction bands, i.e. the kinetic energy
$\epsilon_{\vec{k},i}^{}$ measured from the Fermi level ranges in $[-D,D]$ and
the density of states is $\rho_0=1/(2D)$. We assume $\vec{k}$-independent and symmetric 
tunneling amplitudes $t$ between the dot and both superconducting leads. The dot
has a level energy $\epsilon_d$ and Coulomb interaction $U$. Experimentally, the
crucial characteristic energy scales, namely Coulomb interaction $U$, 
total hybridization $\Gamma=2\pi t^2\rho_0$ and gap $\Delta$, are typically all of the same 
order of magnitude, \cite{wernsdorfersquid, 0_pi_Jorgensen-flensberg} providing a
challenge for analytical methods.

The physics of the quantum dot can be described via its Green's function
\begin{equation}
\label{eq:greensfunction}
\widehat{G}_{d,d}^{}(\tau) = -{\langle}T_{\tau}\Psi_{d}^{}(\tau)\Psi_{d}^{\dagger}(0)\rangle \text{ ,}
\end{equation}
where the Nambu spinor
$$
\Psi_{d}^{}(\tau) = \begin{pmatrix} d_{\uparrow}^{}(\tau) \\ d_{\downarrow}^{\dagger}(\tau) \end{pmatrix}
$$
has been introduced. Because we will only be interested in stationary
equilibrium physics, $\widehat{G}_{d,d}^{}(\tau)$ shall be computed in the
Matsubara frequency formalism.

\subsection{\label{subsec:local_eff_ham}Effective local Hamiltonian}

As the above Hamiltonian has no exact solution, some approximations must be
made. Among the physical ingredients we want to include in a non-perturbative
way is the local pairing on the dot that is crucial for the evolution of the
Andreev bound states. 
Furthermore, the Coulomb interaction shall be taken into account in an exact 
manner in order to describe the atomic states faithfully, and to highlight how 
these are adiabatically connected to the ABS. 
However, the usual development in weak tunnel coupling $t$ around the atomic 
limit \cite{pert_tunnel_glazman-matveev} is not sufficient to describe the
proximity effect at lowest order. Therefore, we shall consider in what follows 
an expansion around a {\it superconducting atomic limit}. 

Such simple solvable limiting case of the model~(\ref{eq:Hamiltonian_complete_normal}) 
is often referred to as the limit of large gap $\Delta \to \infty$, and has been
discussed previously~\cite{H_eff_original_affleck,
Delta_inf_assym_hewson,NRG_spectral_Bauer,vecino_yeyati}. Expansions for finite
$\Delta$ have not however been discussed to our knowledge, and are the topic of
this paper.
We emphasize from the outset (see equation~(\ref{eq:greens_leads_nambu}) below), that 
the superconducting atomic limit as used normally in the literature
corresponds to the limit $D \to \infty$ (i.e. infinite electronic bandwidth),
taken before $\Delta \to \infty$. The order of the two limits is crucial: if the limit
$\Delta \to \infty$ was to be taken first, the dot would be completely decoupled
from the leads and the proximity effect would be lost, so that the limit of infinite 
gap would reduce to the usual atomic limit.
As will be shown now, the superconducting atomic limit should rather be interpreted as 
a low frequency expansion, i.e. a limit where the gap is much larger than the characteristic
frequencies of the dot.

We start off by deriving the Green's function defined in Eq.
\eqref{eq:greensfunction} using the equations of motion. Thereby, the Coulomb
interaction $U$ will at first be omitted for the sake of clarity. Note that in
the end, $U$ will simply give an extra contribution which adds to the
effective Hamiltonian. Fourier transformation straightforwardly yields

\begin{equation}
\label{eq:greens_nambu}
{\widehat{G}_{d,d}}^{-1}(i\omega_n)= i\omega_n\mathds{1}-\epsilon_d
\hat{\sigma}_z 
-t^2\sum_{\vec{k},i}\hat{\sigma}_x
\widehat{G}_{\vec{k}i,\vec{k}i}^{0}(i\omega_n)\hat{\sigma}_x \text{ .}
\end{equation}
In Eq. \eqref{eq:greens_nambu}, $\omega_n$ is a fermionic Matsubara frequency,
$\widehat{G}_{\vec{k}i,\vec{k}i}^{0}(i\omega_n)$ the bare Green's function
in Nambu space of electrons with a wave vector $\vec{k}$ in the lead $i$, and
the Pauli matrices $\hat{\sigma}_\alpha$ have been introduced. Transforming 
the sum over wave vectors $\vec{k}$ into an integral over energies yields

\begin{widetext}
\begin{equation}
\label{eq:greens_leads_nambu}
\sum_{\vec{k}}{\widehat{G}_{\vec{k}i,\vec{k}i}^{0}(i\omega_n)} = 2\rho_0 \arctan\left(\frac{D}{\sqrt{{\omega_n}^2+\Delta_{}^2}}\right)\frac{1}{\sqrt{{\omega_n}^2+\Delta_{}^2}}\begin{pmatrix} -i\omega_n & \Delta \, e^{i\varphi_i} \\ \Delta \, e^{-i\varphi_i} & -i\omega_n \end{pmatrix} \text{ .}
\end{equation} 
\end{widetext}

In the limit $\omega_n \ll \Delta$, the Green's function \eqref{eq:greens_leads_nambu} 
becomes purely {\it static} and reduces to

\begin{equation}
\label{eq:greens_limit}
\sum_{\vec{k}}{\widehat{G}_{\vec{k}i,\vec{k}i}^{0}(i\omega_n)} = 2\rho_0 \arctan\left(\frac{D}{\Delta}\right)\begin{pmatrix} 0 & e^{i\varphi_i} \\ e^{-i\varphi_i} & 0 \end{pmatrix} \text{ .}
\end{equation}
Note that the low frequency limit we consider here yields a Green's function
that indeed depends on the finite bandwidth $D$, and this shows that the limit
$\Delta\to\infty$ shall not be taken for the proximity effect to survive.
In what follows, we will therefore keep both $D$ and $\Delta$ finite.
Plugging Eq. \eqref{eq:greens_limit} into the Green's function
$\widehat{G}_{d,d}^{}(i\omega_n)$ leads to the same result as would have been
obtained with the effective local Hamiltonian

\begin{equation}
\label{eq:H_eff_no_int}
H_\mathrm{eff}^{0} = \sum_\sigma{\epsilon_d^{} \, d_{\sigma}^{\dagger} d_{\sigma}^{}} - \left( \Gamma_{\varphi}\,e^{i\frac{\varphi_L+\varphi_R}{2}} d_{\uparrow}^{\dagger} d_{\downarrow}^{\dagger} + \mathrm{h.c.}\right) \text{ ,}
\end{equation}
where the local pairing amplitude induced by the leads on the dot reads
\begin{equation}
\label{eq:gamma_phi}
\Gamma_{\varphi} = \Gamma \frac{2}{\pi}
\arctan\left(\frac{D}{\Delta}\right)\cos\left(\frac{\varphi}{2}\right).
\end{equation}
which explicitly depends on the ratio $D/\Delta$.
By an appropriate gauge transformation for the operators $d_{\sigma}$, it is
always possible to choose $\Gamma_{\varphi}\,e^{i\frac{\varphi_L+\varphi_R}{2}}
= \left|\Gamma_{\varphi}\right|$, as shall be done from now on. The complete
local effective Hamiltonian is obtained when the Coulomb interaction is taken into
account again. Defining $\xi_d = \epsilon_d + \frac{U}{2}$, the energy level of the
dot is shifted such that the Hamiltonian clearly exhibits particle-hole
symmetry for $\xi_d=0$:

\begin{eqnarray}
\label{eq:H_eff}
H_\mathrm{eff}^{} &=& \sum_\sigma{\xi_d^{} \, d_{\sigma}^{\dagger} d_{\sigma}^{}} - \left|\Gamma_{\varphi}\right|\left(d_{\uparrow}^{\dagger} d_{\downarrow}^{\dagger} + \mathrm{h.c.}\right)\nonumber\\
&&+ \frac{U}{2}\left(\sum_{\sigma}{d_{\sigma}^{\dagger} d_{\sigma}^{}} - 1 \right)^2 \text{ .}
\end{eqnarray}

The physical interpretation of this effective local Hamiltonian is simple. For
finite gap, the quantum dot is coupled to both the Cooper pairs and the
quasiparticles in the leads. The Cooper pairs, which lie at the Fermi level, are
responsible for the proximity effect. The quasiparticles give rise to 
conduction electrons excitations with energies higher than the gap $\Delta$. In the
limit $\omega_n \ll \Delta$, the quasiparticles are far in energy and the
coupling between them and the dot vanishes, which greatly simplifies the physics
and makes an exact solution possible. Yet, as the dot is still coupled to the
Cooper pairs at the Fermi level, the proximity effect survives with a local
pairing term proportional to the hybridization $\Gamma$ between dot
and leads.


\subsection{\label{subsec:spectrum}Spectrum of the effective local Hamiltonian}
As the Coulomb interaction simply yields an extra energy shift of $U/2$ for both
empty and doubly occupied dot, the eigenvectors and eigenvalues of the local
effective Hamiltonian~(\ref{eq:H_eff}) are readily obtained by a Bogoliubov 
transformation~\cite{NRG_spectral_Bauer}, in perfect analogy with solution of 
the BCS Hamiltonian. $H_\mathrm{eff}$ has thus four eigenstates, the singly occupied 
spin $1/2$ states $|\uparrow\rangle$ and $|\downarrow\rangle$ with energy 
$E_{\uparrow}^0=E_{\downarrow}^0=\xi_d$, and two BCS-like states given by

\begin{eqnarray}
|+\rangle &=& u_{}^{} |\uparrow\downarrow\rangle + v_{}^{*} |0\rangle \nonumber \\
\label{eq:states}
|-\rangle &=& -v_{}^{*} |\uparrow\downarrow\rangle + u_{}^{} |0\rangle \text{ ,}
\end{eqnarray}
where $|0\rangle$ denotes the empty dot and $|\uparrow\downarrow\rangle$ the
doubly occupied dot. The amplitudes $u$ and $v$ can always be chosen to be real
with $u = 1/2\,\sqrt{1+\xi_d / \sqrt{{\xi_d}^2+{\Gamma_{\varphi}}^2}}$ and $v =
1/2\,\sqrt{1-\xi_d / \sqrt{{\xi_d}^2+{\Gamma_{\varphi}}^2}}$. The energies
corresponding to these BCS-like states are 
$E_{\pm}^0=U/2 \pm \sqrt{{\xi_d}^2+{\Gamma_{\varphi}}^2}+\xi_d$.

\begin{figure}
		\includegraphics[scale=0.25]{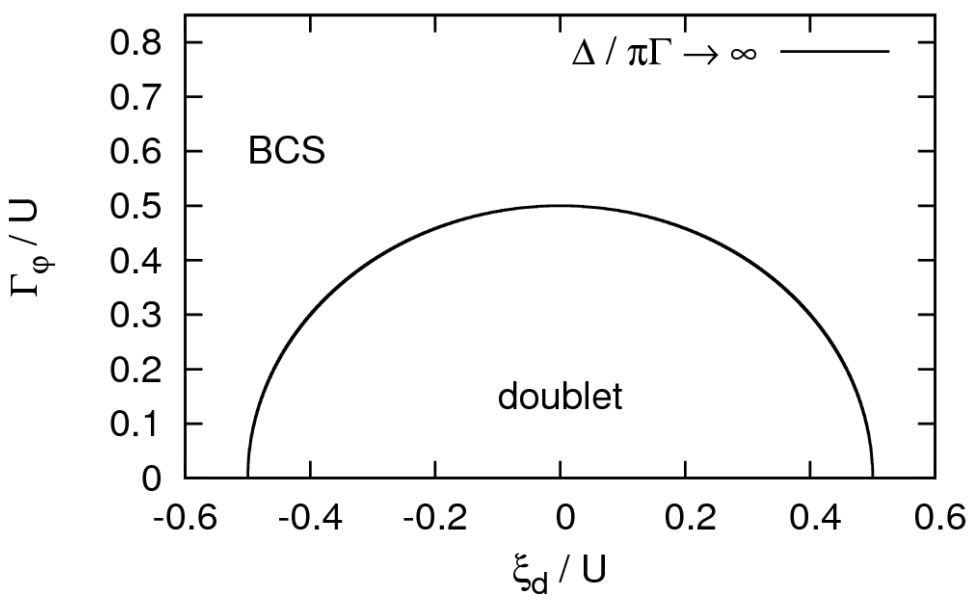}
\caption{Phase diagram of a simple dot with Coulomb interaction $U$, energy
level $\xi_d$ and hybridization $\Gamma$ to superconducting electrodes in the
effective local limit. The transition line corresponds to $E_{\sigma}^0 = E_-^0$.} 
\label{fig:phase-diag-inf}
\end{figure}
As $E_+^0$ is always larger than $E_-^0$, the effective local Hamiltonian has two
possible ground states: the low energy BCS-like state $|-\rangle$ or the degenerate spin
$1/2$ doublet $\{|\uparrow\rangle,|\downarrow\rangle\}$. In the $|-\rangle$
state, the energy is minimized for $\varphi = 0$. Thus, the spin singlet phase
corresponds to a $0$-junction (a result well known from the weak coupling
limit~\cite{pert_tunnel_glazman-matveev}). The transition between the singlet phase 
and the spin $1/2$ doublet takes place at ${\xi_d}^2+\Gamma_\varphi^2 = U^2/4$,
and Fig. \ref{fig:phase-diag-inf} shows the corresponding phase 
diagram for variable $\xi_d$, $\Gamma_\varphi$ and $U$. The state adopted by the quantum 
dot in the large gap limit therefore results from a competition between the local pairing 
(induced by the proximity effect and characterized by the hybridization $\Gamma$) and 
the Coulomb interaction.

\subsection{\label{subsec:ABS}Andreev bound states}

As outlined in the introduction, the coupling to superconducting leads induces a
gap in the spectral function of the dot, inside which discrete Andreev bound 
states can form.
The spectral function of the dot shows therefore sharp peaks, which could be measured 
by STM \cite{Governale1} or microwave-\cite{spectrum_fogelstrom}optical 
\cite{josephson_diode} experiments as proposed recently.
These peaks indicate addition energies at which an electron may enter (or leave) the 
dot, and correspond therefore to transitions between states with $n$
and $n\pm1$ electrons. Hence, the ABS peaks in the spectral function may be interpreted 
as transitions between the superconducting atomic levels of the dot
$\{|\sigma\rangle,|+\rangle,|-\rangle\}$, possibly renormalized by single-particle 
tunneling events neglected in $H_\mathrm{eff}$ (to be included in the next section).
Furthermore, transitions from a spin $1/2$ doublet to a spin singlet
necessarily involve an electron exchange between the dot and the superconducting
leads. As the states $|-\rangle$ and $|+\rangle$ correspond to the superposition
of an empty and doubly occupied dot, this electron exchange and the final
singlet states can be understood within the Andreev reflection picture.

Putting everything together, our effective local Hamiltonian in Eq.
\eqref{eq:H_eff} describes the energies of the Andreev bound states as
transition energies from the spin 1/2 doublet to the spin singlet
states.\cite{NRG_spectral_Bauer, Delta_inf_assym_hewson} There are thus four
Andreev bound states in the large gap limit for the model
(\ref{eq:Hamiltonian_complete_normal}), with energy $\pm a_0$ and $\pm b_0$ which read:

\begin{eqnarray}
\label{eq:a0}
a_0 &=& E_-^0 - E_{\sigma}^0 = \frac{U}{2} - \sqrt{{\xi_d}^2+{\Gamma_{\varphi}}^2}\\
\label{eq:b0}
b_0 &=& E_+^0 - E_{\sigma}^0 = \frac{U}{2} + \sqrt{{\xi_d}^2+{\Gamma_{\varphi}}^2} \text{ .}
\end{eqnarray}
The $0$/$\pi$ transition corresponds to the crossing of the $|-\rangle$ and
$|\sigma\rangle$ states, which occurs for $a_0 = 0$.

\section{\label{sec:pert}Perturbation expansion around the effective local Hamiltonian}
\subsection{\label{subsec:pert}Perturbation theory}
The effective Hamiltonian is not sufficient to obtain satisfying results
for all regimes of parameters. First, $H_\mathrm{eff}$ only describes the
$0$/$\pi$-junction transition due to the competition between a local moment
state (stabilized by the Coulomb blockade) and a spin singlet (induced by the
proximity effect). However, if the Coulomb interaction is strong (i.e. $U
\gg \Gamma, |\xi_d|$ below the Kondo temperature), the local moment can be screened by the Kondo
effect, which will compete with the superconducting gap for the $0-\pi$ transition, so that a typical scaling in the ratio of the Kondo temperature to
the gap $\Delta$ will appear. Also, the Josephson current in the $\pi$-phase
identically vanishes from $H_\mathrm{eff}$, as the spin doublet does not disperse
with the superconducting phase difference, a limitation of the large gap limit.
On a more quantitative basis, the experimental gap $\Delta$ is usually of the order of a 
few kelvins, which is also the typical scale for both $\Gamma$ and $U$ in
carbon nanotube quantum dot devices.

In order to extend the description of the quantum dot's physics, energy
corrections shall be calculated with a perturbation theory around the effective
Hamiltonian \eqref{eq:H_eff}. Once these corrections have been obtained,
physical observables like the Josephson current may be computed via the free
energy $F = -\frac{1}{\beta} \ln(Z)$, with $\beta$ the inverse temperature. 
Therefore, it is most convenient to work in an action based description, which 
directly yields the partition function $Z$. Following Ref. \onlinecite{meanf_arovas}, 
we first integrate over the fermions in the leads. Omitting the resulting irrelevant 
constant, the partition function reads

\begin{eqnarray}
\label{eq:Z_effective_action}
Z &=& \int{\mathcal{D}(\overline{\Psi}_d,\Psi_d)e^{-S_\mathrm{dot}}} \text{ with}\\
S_\mathrm{dot} &=& \sum_{\vec{k},i,\omega_n}{\overline{\Psi}_{d,n} \widehat{H}_{T_i}^{} \widehat{G}_{\vec{k}i,\vec{k}i}^{0}(i\omega_n) \widehat{H}_{T_i}^{\dagger} \Psi_{d,n}}\nonumber\\
&&+ \sum_{\omega_n}{\overline{\Psi}_{d,n} \begin{pmatrix} -i\omega_n + \epsilon_d & 0 \\ 0 & -i\omega_n - \epsilon_d \end{pmatrix} \Psi_{d,n}} \nonumber\\
&&+ \int_{0}^{\beta}{d\tau \, U \overline{d}_{\uparrow}(\tau)\overline{d}_{\downarrow}(\tau)d_{\downarrow}(\tau)d_{\uparrow}(\tau)} \label{eq:action_dot} \text{ ,}
\end{eqnarray}
where we have introduced the Grassmann Nambu spinors at Matsubara frequency
$\omega_n=(2n+1)\pi/\beta$,

\begin{eqnarray*}
\Psi_{d,n} &=& \frac{1}{\sqrt{\beta}}\sum_{\omega_n}{\begin{pmatrix} d_{\uparrow}^{}(\tau) \\ \overline{d}_{\downarrow}^{}(\tau) \end{pmatrix}e^{-i\omega_n\tau}} \text{ and }\\
\overline{\Psi}_{d,n} &=& \frac{1}{\sqrt{\beta}}\sum_{\omega_n}{\begin{pmatrix} \overline{d}_{\uparrow}^{}(\tau), & d_{\downarrow}^{}(\tau) \end{pmatrix}e^{i\omega_n\tau}} \text{ ,}
\end{eqnarray*}
denoting the Grassmann fields associated with electrons in the dot by $\overline{d}_{\sigma}$ and $d_{\sigma}$.

The perturbation consists of the terms in Eq. \eqref{eq:Z_effective_action} that
are not contained in the action $S_\mathrm{eff}$ corresponding to the effective
local Hamiltonian. A simple identification yields

\begin{widetext}
\begin{eqnarray}
\label{eq:S_eff}
S_\mathrm{eff} &=& \int_0^{\beta}{d\tau \left( \sum_{\sigma} \overline{d}_\sigma(\tau) (\frac{\partial}{\partial\tau} + \epsilon_d) d_{\sigma}(\tau) - |\Gamma_{\varphi}^{}|\overline{d}_{\uparrow}(\tau)\overline{d}_{\downarrow}(\tau) - |\Gamma_{\varphi}^{}|d_{\downarrow}(\tau)d_{\uparrow}(\tau) + U \overline{d}_{\uparrow}(\tau)\overline{d}_{\downarrow}(\tau)d_{\downarrow}(\tau)d_{\uparrow}(\tau)\right)} \text{ ,}\\
\label{eq:S_pert}
S_\mathrm{pert} &=& \int_0^{\beta}{d\tau \int_0^{\beta}{d\tau' \sum_{\vec{k},i}{ \overline{\Psi}_d(\tau) \widehat{H}_{T_i}^{} \widehat{G}_{\vec{k}i,\vec{k}i}^0(\tau-\tau') \widehat{H}_{T_i}^{\dagger} \Psi_d(\tau')}}} + \int_{0}^{\beta}{d\tau \left( |\Gamma_{\varphi}^{}|\overline{d}_{\uparrow}(\tau)\overline{d}_{\downarrow}(\tau) + |\Gamma_{\varphi}^{}|d_{\downarrow}(\tau)d_{\uparrow}(\tau)\right)} \text{ .}
\end{eqnarray}
\end{widetext}
Note that $S_\mathrm{eff}$ contains the local pairing term derived in section
\ref{subsec:local_eff_ham}. The proximity effect is thus treated
non-perturbatively (just like the Coulomb interaction), which is the crucial
ingredient of our analytic approach. The perturbation $S_\mathrm{pert}$ simply
corresponds to the tunnel coupling between the dot and the electrodes other than
the lowest order proximity effect.

The actual corrections are calculated by expanding the partition function to the
first order in $S_\mathrm{pert}$ according to

\begin{eqnarray}
Z &=& \int{\mathcal{D}(\overline{\Psi},\Psi)e^{-S_\mathrm{eff}-S_\mathrm{pert}}}\nonumber \\
&\approx&
\int{\mathcal{D}(\overline{\Psi},\Psi)e^{-S_\mathrm{eff}}\left(1-S_\mathrm{pert}+\ldots\right)}
\end{eqnarray}
which we then identify with
\begin{eqnarray}
Z &=&\sum_{\sigma}{e^{-\beta E_{\sigma}}}+ e^{-\beta E_+} + e^{-\beta E_-} \text{ ,}
\end{eqnarray}
where the renormalized superconducting atomic levels 
$E_\sigma = E_{\sigma}^0+\delta E_{\sigma}$ and
$E_\pm = E_{\pm}^0+\delta E_{\pm}$ are obtained from:
\begin{eqnarray}
e^{-\beta E_{\sigma}} &\approx& e^{-\beta E_{\sigma}^0}\left(1-\beta \delta E_{\sigma}\right) \\
e^{-\beta E_{\pm}} &\approx& e^{-\beta E_{\pm}^0}\left(1-\beta \delta E_{\pm}\right) \text{ .}
\end{eqnarray}

Because the Coulomb interaction is taken into account, Wick's theorem cannot be
used to calculate $Z$. Instead, expectation values are calculated using Lehmann
representation. Explicit calculations may be found in the appendix. In the zero
temperature limit $\beta \to \infty$, the energy corrections are

\begin{widetext}
\begin{eqnarray}
\nonumber
\delta E_\sigma \!&=&\! -t^2\sum_{\vec{k}}{\left(\frac{1}{E_{\vec{k}} + (E_+^0-E_{\sigma}^0)} + \frac{1}{E_{\vec{k}} + (E_-^0-E_{\sigma}^0)} + \frac{2\Delta}{E_{\vec{k}}}uv\left|\cos(\frac{\varphi}{2})\right|\left(\frac{1}{E_{\vec{k}} + (E_+^0-E_{\sigma}^0)} - \frac{1}{E_{\vec{k}} + (E_-^0-E_{\sigma}^0)}\right)\right)}\\
\label{eq:Es}\\
\label{eq:E+}
\delta E_+ \!&=&\! -t^2\sum_{\vec{k},\sigma}{\left(\frac{1}{E_{\vec{k}}-(E_+^0-E_{\sigma}^0)}-\frac{2\Delta}{E_{\vec{k}}}uv\left|\cos(\frac{\varphi}{2})\right|\frac{1}{E_{\vec{k}}-(E_+^0-E_{\sigma}^0)}\right)}-2|\Gamma_{\varphi}|uv\\
\label{eq:E-}
\delta E_- \!&=&\! -t^2\sum_{\vec{k},\sigma}{\left(\frac{1}{E_{\vec{k}}-(E_-^0-E_{\sigma}^0)}+\frac{2\Delta}{E_{\vec{k}}}uv\left|\cos(\frac{\varphi}{2})\right|\frac{1}{E_{\vec{k}}-(E_-^0-E_{\sigma}^0)}\right)}+2|\Gamma_{\varphi}|uv \text{ ,}
\end{eqnarray}
\end{widetext}
with the quasiparticle energy $E_{\vec{k}} = \sqrt{{\epsilon_{\vec{k}}}^2+{\Delta}^2}$.

\subsection{\label{subsec:renorm}Self-consistent renormalization of the energy}

Eqs. \eqref{eq:Es}-\eqref{eq:E-} yield the first corrections to
the energy levels, so that the bound states energies $a_0$ and $b_0$ are simply shifted by 
$\delta a = \delta E_- - \delta E_{\sigma}$ and $\delta b = \delta E_+ - \delta E_{\sigma}$.
Obviously, these expressions are logarithmically divergent when the bound states
energies $a_0$ and $b_0$ approach the gap edge, and are therefore only valid
as long as e.g. $a_0\gg\Gamma \log[(D+\Delta)/(\Delta-a_0)]$.
In the limit of large gap $\Delta\gg a_0$, these corrections to $a_0$ are thus of
the order $\Gamma a_0/\Delta$, so that the small dimensionless parameter is 
indeed $\Gamma/\Delta$. However, this peculiar logarithmic dependence of the
ABS energy renormalization shows that doing a straightforward $1/\Delta$
expansion around the effective local Hamiltonian will be rapidly uncontrolled,
and will have a hard time reproducing the logarithmic singularities at $\Delta$ 
close to $a_0$. For this reason, and also because the large gap limit becomes
trivial for a finite electronic bandwidth, as discussed in
section~\ref{subsec:local_eff_ham}, it was indeed more appropriate to single out in the
total action all terms left over with respect to the local superconducting effective
Hamiltonian, see equation~(\ref{eq:S_pert}), and do perturbation theory around these.

Because our lowest-order expansion obviously still breaks down when the gap 
becomes comparable to the bound state energy, one would naturally seek to resum the
leadind logarithmic divergences in equations~(\ref{eq:Es})-(\ref{eq:E-}).
This can be achieved using a self-consistency condition inspired by Brillouin-Wigner
perturbation theory, \cite{BW_pert_rama_sur} which allows to extend greatly
the regime of validity of the perturbative scheme. The resulting self-consistent 
equations that we obtain are:

\begin{eqnarray}
\label{eq:a}
&\delta a& = -\frac{\Gamma}{\pi}\int_{0}^{D}{d\epsilon \, \left(\frac{2}{E-a(\Delta)}-\frac{1}{E+b_0}-\frac{1}{E+a_0}\right.}\nonumber\\
&+&\left.\frac{\Delta}{E}uv\left|\cos\left(\frac{\varphi}{2}\right)\right|\left(\frac{2}{E-a(\Delta)}-\frac{1}{E+b_0}+\frac{1}{E+a_0}\right)\right)\nonumber\\
&+&2|\Gamma_{\varphi}|uv
\end{eqnarray}
and
\begin{eqnarray}
\label{eq:b}
&\delta b& = -\frac{\Gamma}{\pi}\int_{0}^{D}{d\epsilon \, \left(\frac{2}{E-b(\Delta)}-\frac{1}{E+b_0}-\frac{1}{E+a_0}\right.}\nonumber\\
&+&\left.\frac{\Delta}{E}uv\left|\cos\left(\frac{\varphi}{2}\right)\right|\left(\frac{-2}{E-b(\Delta)}-\frac{1}{E+b_0}+\frac{1}{E+a_0}\right)\right)\nonumber\\
&-&2|\Gamma_{\varphi}|uv \text{ ,}
\end{eqnarray}
with $E=\sqrt{\epsilon^2+\Delta^2}$, and $a_0,b_0$ have been defined in Eqs.
(\ref{eq:a0})-(\ref{eq:b0}), with $a(\Delta) = a_0 + \delta a$,
$b(\Delta) = b_0 + \delta b$. Note that terms like $1/(E+a_0)$ have no
self-consistency because there are no associated divergences. Eq. \eqref{eq:a}
and \eqref{eq:b} clearly now hold as long as the {\it renormalized} energies 
$a(\Delta)$ and $b(\Delta)$ are not too close to the gap edge, $\pm \Delta$ 
respectively.

\section{\label{sec:results}Results}
\subsection{\label{subsec:phase_diag}Phase diagram}
\begin{figure}
\includegraphics[scale=0.25]{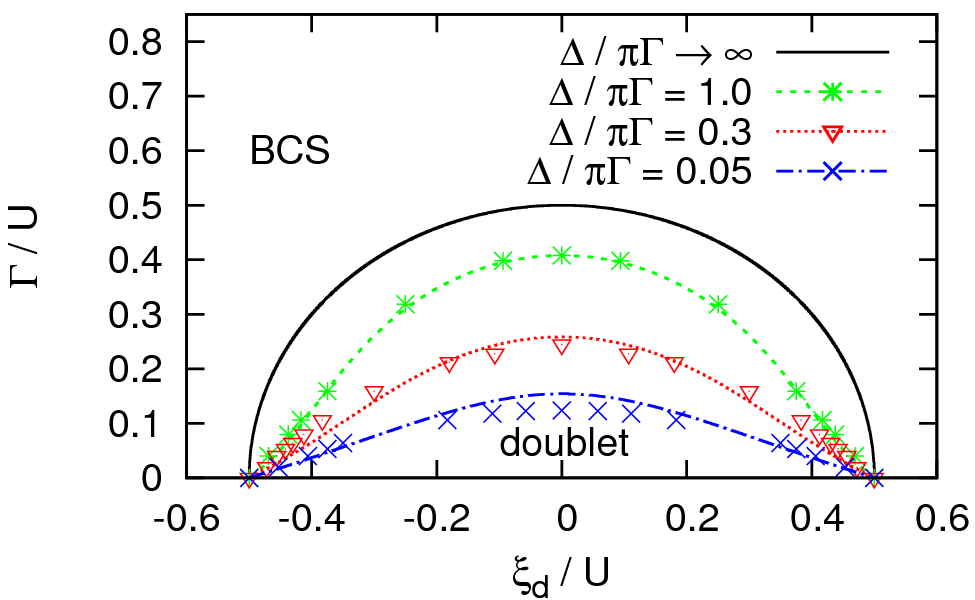}
\caption{(Color online) Phase diagram of a simple dot with Coulomb interaction $U$,
tunnel coupling $\Gamma$ to superconducting electrodes with gap $\Delta$ for
$\varphi = 0$ and $\pi\Gamma = 0.2 \, D$. The symbols indicate NRG data from
Ref. \onlinecite{NRG_spectral_Bauer} and the various lines our results.}
\label{fig:phase-diag}
\end{figure}

We start by discussing the $0-\pi$ transition line, by comparison to
the numerical renormalization group (NRG) data by Bauer 
et al.\cite{NRG_spectral_Bauer}. Fig. \ref{fig:phase-diag} shows the 
extension to smaller gaps $\Delta$ values of the phase diagram obtained 
with unrenormalized local superconducting states for infinite gap 
(Fig. \ref{fig:phase-diag-inf}).
Even though our perturbative approach is fairly simple, the results reproduce
nicely the NRG data of Refs. \onlinecite{NRG_spectral_Bauer} and
\onlinecite{NRG_current_Karrasch}. The analytically obtained phase diagram is
indeed identical to the NRG data for $\Delta \gtrsim \Gamma$. For
smaller $\Delta/\Gamma$, the Kondo effect sets in, but the transition lines remain 
quantitatively correct for $\xi_d$ near $\pm U/2$, with increasing deviations 
from the NRG calculations close to the particle-hole symmetric point $\xi_d=0$ at 
large Coulomb interaction U.
In this regime, the $0$-phase possesses a Kondo singlet ground state. As the
leads are superconductors, the formation of a Kondo resonance involves the
breaking of Cooper pairs. Therefore, the transition is now due to the
competition between $T_K$ and the superconducting gap $\Delta$, and should occur
at $k_B \, T_K \propto \Delta$.

\begin{figure}
\centering
\includegraphics[scale=0.25]{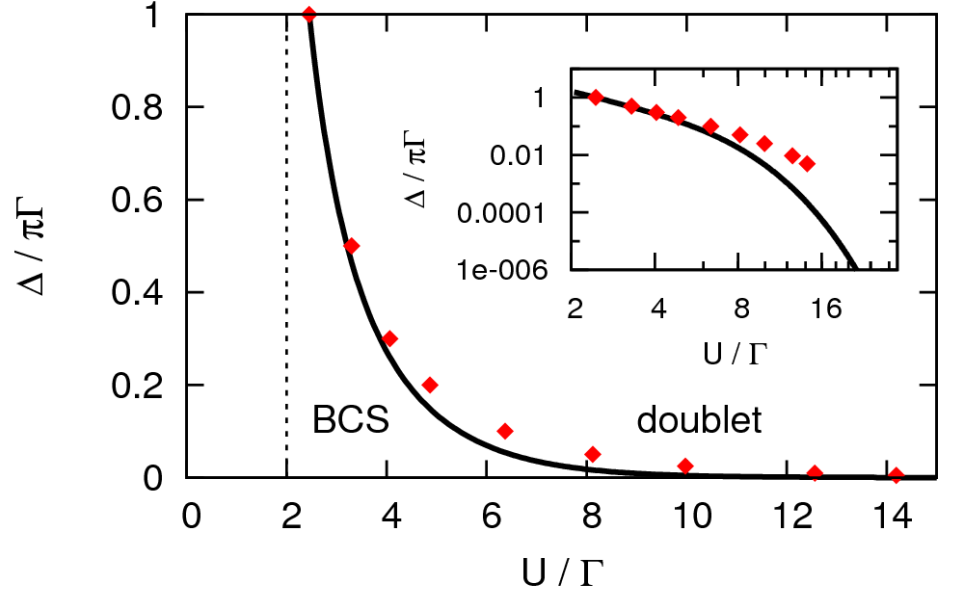}
\caption{(Color online) Transition line between a doublet state and the BCS-like state
at particle hole symmetry $\xi_d = 0$ (solid curve) for $\varphi = 0$ and
$\pi\Gamma = 0.2 \, D$. The vertical dotted line corresponds to the transition
asymptote in the effective local limit at $\Delta \to \infty$. The dots indicate
NRG data from Ref. \onlinecite{NRG_spectral_Bauer} and the solid line our result. The inset 
displays the same curves on a logarithmic scale.}
\label{fig:TK}
\end{figure}

Fig. \ref{fig:TK} shows a plot of the transition line for $\xi_d = 0$ as
obtained with Eq. \eqref{eq:a} (solid curve). The vertical, dotted line depicts
the asymptote in the effective local limit. The symbols again correspond to NRG
data.\cite{NRG_spectral_Bauer} The Kondo temperature is given by $T_K = 0.182 \,
U \sqrt{8\Gamma/\pi U}e^{-\pi U/8\Gamma}$ (see for example Ref.
\onlinecite{NRG_spectral_Bauer}). The inset shows on a log-log scale 
that our approach captures an exponential
decay of the transition line with the Coulomb interaction. Nonetheless, the
suppression of the BCS-like phase appears quantitatively stronger than
expected: a factor 4 instead of 8 is found in the exponential factor of $T_K$. 
The reason for this is that the vertex renormalizations have not been 
taken into account, as discussed in the context of U-NCA \cite{UNCA}.
Far away from the particle-hole symmetric limit, our results for the Kondo
temperature reproduce the lowest-order scaling theory for the infinite-$U$
Anderson model \cite{Haldane}, and are in relatively good agreement with 
NRG data for all $\Delta/\Gamma$ values.

\subsection{\label{subsec:p_h_s}Energy renormalizations at particle hole symmetry ($\xi_d=0$) }

\begin{figure}
\includegraphics[scale=0.30]{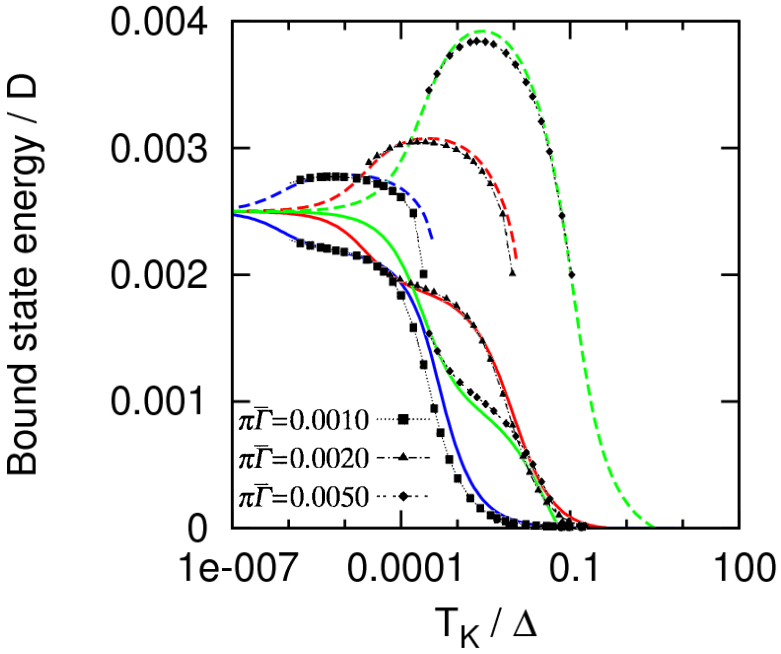}
\caption{(Color online) Renormalization of the Andreev bound state energies as a
function of $T_K/\Delta$ (the Kondo temperature is given in the text). The
dashed curves correspond to the high energy bound state $b(\Delta)$, the solid
curves correspond to $a(\Delta)$. All curves have been calculated for $U = 0.005
\, D$ and $\xi_d = 0$, with several hybridization values $\pi\Gamma/D =
0.001,0.002,0.005$ (from left to right). Symbols are the NRG results obtained 
in Ref. \onlinecite{NRG_energy_Ohashi}.}
\label{fig:scaling-D}
\end{figure}

\begin{figure}
\includegraphics[scale=0.30]{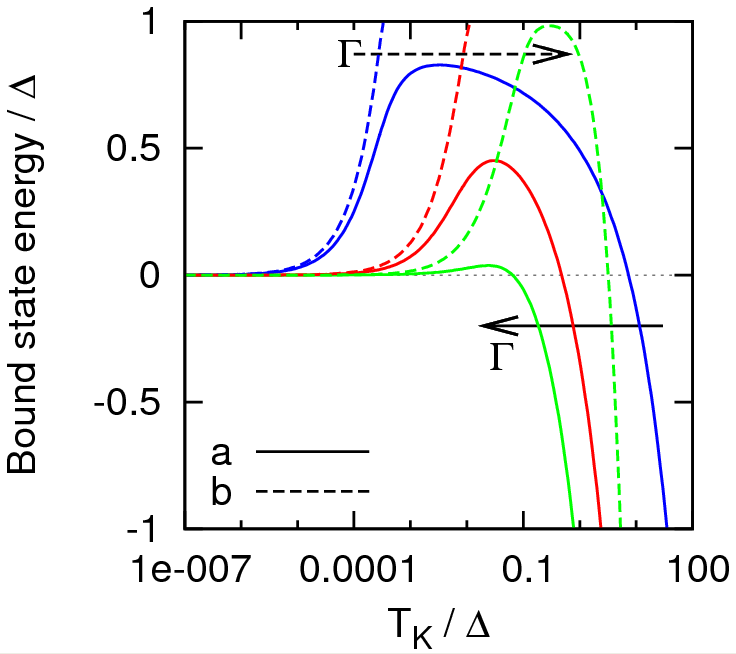}
\caption{(Color online) Same data as in Fig. \ref{fig:scaling-D}, but normalized by the gap.}
\label{fig:scaling-delta}
\end{figure}

While Fig. \ref{fig:phase-diag} only indicates the transition line between the
spin $1/2$ doublet and the lowest BCS spin singlet, it is also instructive to 
look at the actual renormalization of the energy levels while varying the gap $\Delta$ from
large values to smaller ones beyond the critical point. 
Fig. \ref{fig:scaling-D} indicates the renormalized energies of the two Andreev
bound states (i.e. the difference between the spin $1/2$ doublet and the two
spin singlets energies) for different hybridizations $\Gamma$. 
We note that our results are in quantitative agreement with the NRG
calculations of Yoshioka and Ohashi.\cite{NRG_energy_Ohashi} 
Several regions need to be distinguished.
If the gap $\Delta$ is much larger than the bandwidth $D$, all curves collapse
at the value $U/2$ (left hand side of Fig. \ref{fig:scaling-D}), since there is no
hybridization with both quasiparticles and Cooper pairs anymore, and one
recovers the bare atomic levels.
When the gap starts to decrease, the proximity effect simply splits the two Andreev 
bound states according to equations (\ref{eq:a0})-(\ref{eq:b0}).
When the gap becomes of the same order as the typical energy scales of the dot
$a_0$ and $b_0$, the superconducting atomic levels start to mix with the
electrodes, so that the energies renormalize in a non trivial way.
One can see that the transition involving the highest BCS states ends up
touching the gap edge for $\Delta \approx U/2$, so that half of the ABS are
absorbed into the continuum above $\Delta$, as can be seen in Fig. \ref{fig:scaling-delta}.
The lowest BCS state follows however a downward renormalization, until the Fermi level is
crossed and the ground state becomes the $0$-state.
The difference in behavior between the lowest and highest bound states (the 
former being never allowed to leave the superconducting gap) can be tracked 
into equations (\ref{eq:a})-(\ref{eq:b}), where level repulsion effects from 
the gap edge occur for the low energy level $|-\rangle$ but are canceled for
the high energy level $|+\rangle$, which is hence allowed to penetrate
into the continuum. These considerations unveil how the ABS may be adiabatically
connected to the atomic (or molecular) levels in a complicated fashion.

Again, our simple analytic approach reproduces the NRG results \cite{NRG_energy_Ohashi} 
over a vast regime of parameters. Yet, some deviations are observed in the Kondo regime: we find (for the
highest hybridization $\pi\Gamma = 0.005 \, D$) that the high energy BCS-like
state is not absorbed anymore into the continuum of states - an artifact of the limits
of our perturbative approach. Notice also that the energy corrections are too
important if the gap becomes very small, an effect actually due to our underestimation 
of the Kondo temperature at particle-symmetry, as discussed previously. Finally 
Fig. \ref{fig:scaling-delta} shows that, in the limit of vanishing gap, our approach 
is only valid as long as $a(\Delta) \geq -\Delta$ (as has been mentioned in section
\ref{subsec:renorm}), because the lowest bound state artificially escapes from
the gap. The expected saturation of $a(\Delta)$ near $-\Delta$ can be restored 
by adding a further self-consistency for terms such as $1/(E+a_0)$ in (\ref{eq:a})
(not shown here).

\subsection{\label{subsec:off_p_h_s}Energy renormalizations outside particle hole symmetry ($\xi_d \ne 0$)}
From an experimental point of view, the position of the energy level of the
quantum dot is the most controllable parameter of the system (by a
simple gate voltage). Therefore, it is important to analyze the evolution 
of the Andreev bound states for different values of $\xi_d$.

\begin{figure}
\includegraphics[scale=0.30]{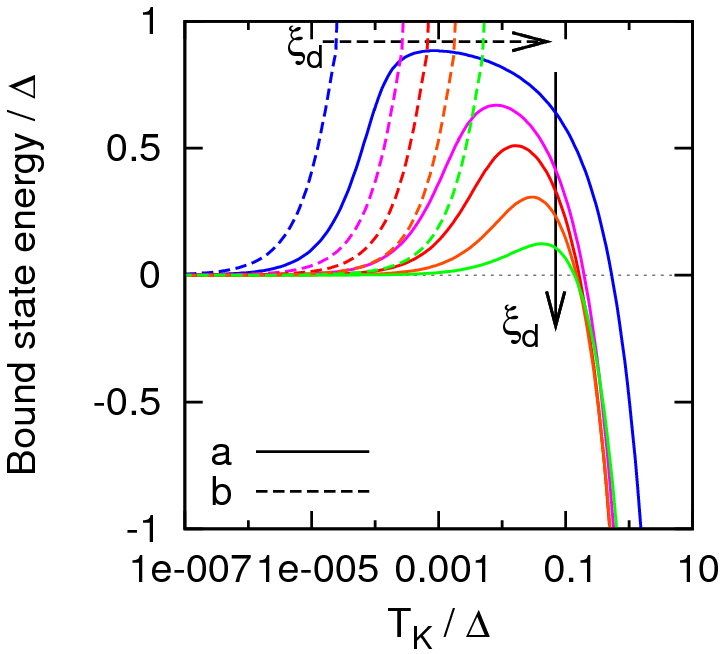}
\caption{(Color online) Renormalization of the Andreev bound state energies outside particle-hole 
symmetry. The dotted curves correspond to the high energy bound state
$b(\Delta)$, the solid curves correspond to $a(\Delta)$. All curves have
been calculated for $U = 0.5 \, D$ and $\pi\Gamma = 0.05 \, D$, with several
level shifts $\xi_d/U = 0.3, 0.375, 0.4, 0.425, 0.45$. Quantitatively similar
results were obtained by the NRG in Ref. \onlinecite{NRG_energy_Ohashi}}
\label{fig:scaling-off_ph}
\end{figure}

Fig. \ref{fig:scaling-off_ph} illustrates how the energies of the bound states
scale with $\Delta$ for $\xi_d \neq 0$ and can be favorably compared to the NRG data by
Yoshioka and Ohashi.\cite{NRG_energy_Ohashi} The more particle-hole symmetry is
broken, the more the low energy bound state moves away from the gap edge,
ensuring even better convergence of our expansion for a given value of $\Gamma$. 
This can be understood given that this bound state corresponds to the transition between
$|-\rangle$ and the spin $1/2$ doublet: outside particle-hole symmetry, the dot
either seeks to be as empty as possible (for $\xi_d > 0$) or as occupied as
possible (for $\xi_d < 0$). Thus, a BCS-like wave function will be favored. As a
consequence, the Kondo effect (that necessitates a singly occupied dot) 
is less favored. This corresponds to a regime where our approximation scheme
works at best.

\begin{figure}
\includegraphics[scale=0.25]{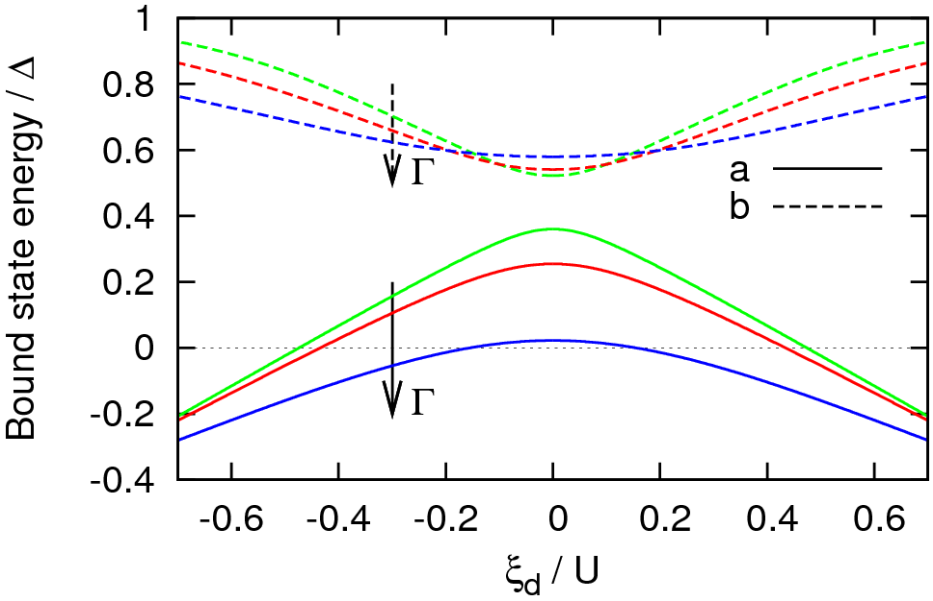}
\caption{(Color online) Evolution of the Andreev bound state energies as a
function of the dot's energy level for $U = 0.005 \, D$ and $\Delta = U$. The hybridization
takes several values $\pi\Gamma/D = 0.001, 0.002, 0.005$.}
\label{fig:scaling-off_ph_2}
\end{figure}

Further understanding can be gained by looking at the energies of the Andreev
bound states as a function of $\xi_d$ on Fig. \ref{fig:scaling-off_ph_2}.
We recover the fact that the high energy bound states increases in energy by
breaking particle-hole symmetry, whereas the low energy bound state has a
decreasing energy. In addition, Fig. \ref{fig:scaling-off_ph_2} shows that the
dispersion of both ABS weakens for increasing hybridization.
Indeed, the more the dot is hybridized with the leads,
the less the Andreev bound state energy is sensitive to 
the bare values of the dot parameters.


\subsection{\label{subsec:correl}Superconducting correlations on the dot}
In order to further analyze the evolution of the states in the dot as a function
of the parameters in the model~(\ref{eq:Hamiltonian_complete_normal}), we
investigate now the superconducting
correlations $\langle d_{\uparrow}^{\dagger} d_{\downarrow}^{\dagger} \rangle$
on the dot. For the effective local Hamiltonian, these correlations are zero in
the spin doublet phase. In the BCS-like phase, the correlations are maximal if
the two states $|0\rangle$ and $|\uparrow\downarrow\rangle$ are equivalent, i.e.
at particle hole symmetry. If the dot level is far from $\xi_d = 0$, the wave
function will be predominantly $|0\rangle$ (if $\xi_d$ is positive) or
$|\uparrow\downarrow\rangle$ (if $\xi_d$ is negative). This obviously kills the
superconducting correlations.

\begin{figure}
\centering
\includegraphics[scale=0.25]{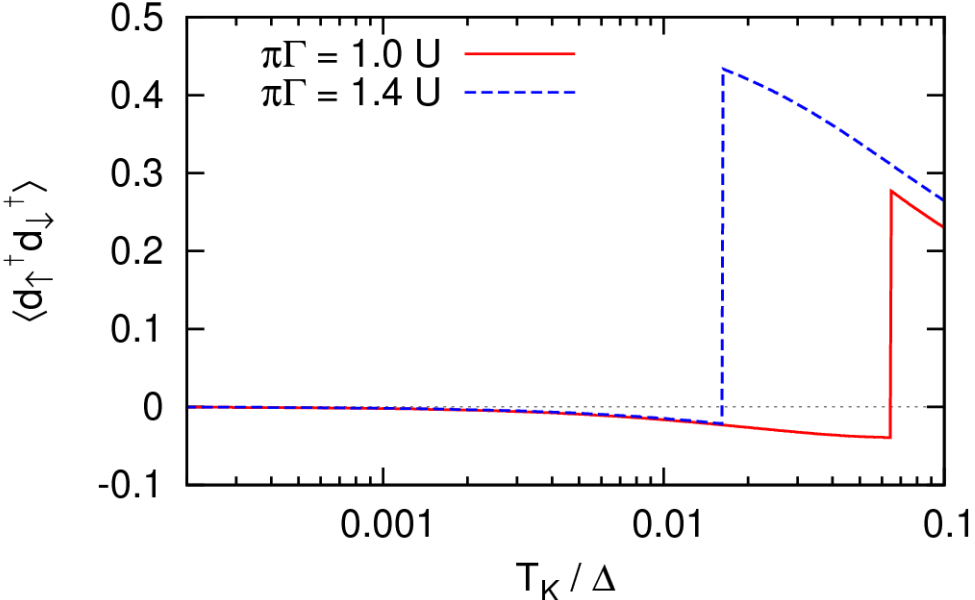}
\caption{(Color online) Superconducting correlations as a function of the gap $\Delta$ 
(for $U=D/200$ and $\xi_d = 0$).}
\label{fig:correl_delta}
\end{figure}

As the gap decreases from infinity, the (formerly) singly occupied state will start 
having a BCS-like admixture and therefore a non zero superconducting correlation.
In contrast, the mixing will result in a decreased correlation in the BCS-like
phase. Nevertheless, if the gap tends to zero, one would expect the correlations
to vanish as well. This is indeed what Fig. \ref{fig:correl_delta} shows. For
large gaps, the dot is in the spin $1/2$ phase; the correlations are small, but
increase as the states mix. The transition to the BCS-like phase results in a
discontinuous jump in the correlations, before they finally vanish for very
small gaps. It can thus be concluded that the correlations should be normalized
by the gap if one is interested in measuring only the mixing effect. Finally,
the two different curves show how hybridization stabilizes the BCS-like state
with respect to the spin doublet via the $0-\pi$ transition.

\begin{figure}
\centering
\includegraphics[scale=0.25]{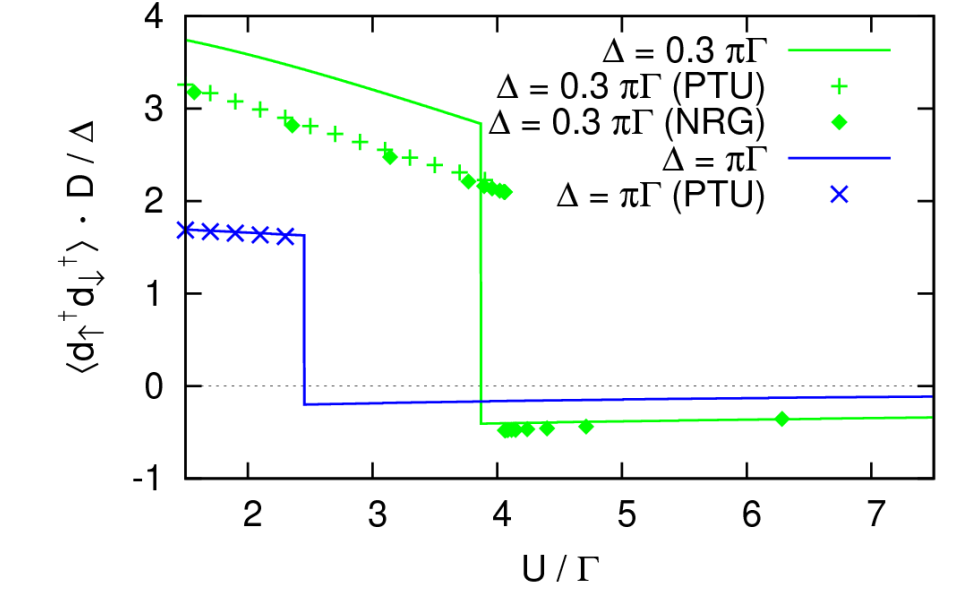}
\caption{(Color online) Superconducting correlations as a function of the Coulomb interaction
$U$, for $\Delta/(\pi\Gamma) = 0.3, 1.0$ and at particle-hole symmetry $\xi_d = 0$. Solid 
lines are the results of our self-consistent equations, 
diamonds correspond to NRG data from Ref. \onlinecite{NRG_spectral_Bauer},
and crosses are perturbation theory in $U$ at second order.}
\label{fig:correl_U}
\end{figure}

As the Coulomb interaction tries to prevent the formation of a Cooper pair wave
function, the transition between the BCS-like phase and the spin doublet can
also be achieved if the Coulomb interaction is tuned, as shown in Fig.
\ref{fig:correl_U}. The effect of the mixing is clearly visible by an increase
of the correlation $\langle d_{\uparrow}^{\dagger} d_{\downarrow}^{\dagger} \rangle$
(now normalized by the gap) while $U$ is lowered. We also find that the correlations 
relative to the gap decrease for higher gaps, which is a simple saturation effect 
(the highest possible correlations are $\langle d_{\uparrow}^{\dagger}
d_{\downarrow}^{\dagger} \rangle = 0.5$). Furthermore, our results are
quantitatively precise if the gap if larger than the hybridization
$\Gamma$ for all values of $U$, while relatively small deviations appear for
$\Delta<\pi \Gamma$, as shown by the comparison to
the NRG data from Ref. \onlinecite{NRG_spectral_Bauer}, and to second-order
perturbation theory in $U$ (valid in the singlet phase only, providing accurate
results for $U\lesssim 2 \pi \Gamma$ roughly) \cite{pert_U_vecino-yeyati,meng}.

\begin{figure}
\centering
\includegraphics[scale=0.25]{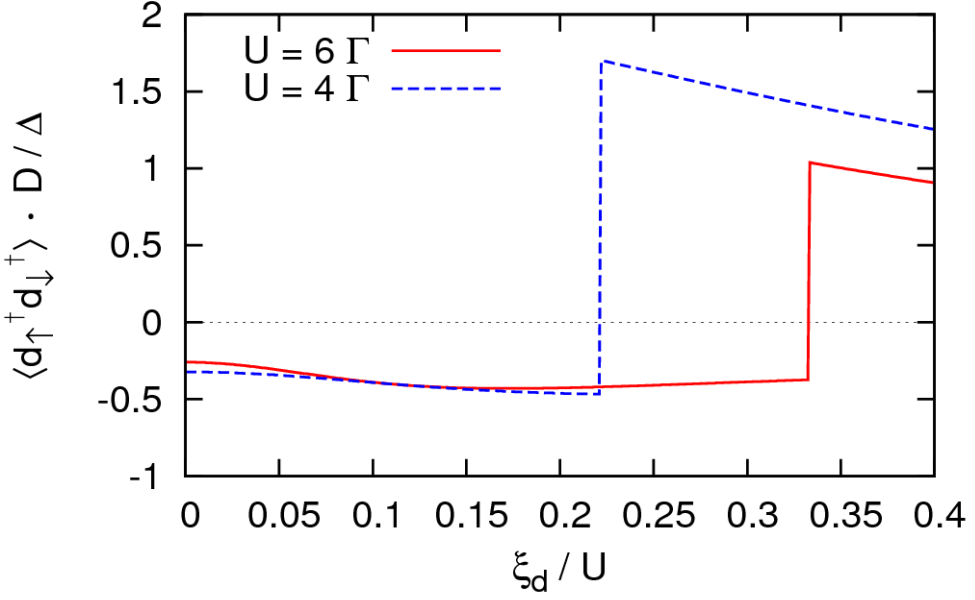}
\caption{(Color online) Superconducting correlations outside particle hole symmetry 
(for $\pi\Gamma = 0.2 \, D$, $U = 6\, \Gamma$ and $\Delta = 0.1 D$). }
\label{fig:correl_xi}
\end{figure}

Finally, we analyze how the correlations evolve outside particle hole symmetry.
As mentioned above, one expects the correlations to decrease because the dot
evolve from a superconducting atomic limit toward a usual atomic
limit (i.e. from the states $|\pm\rangle$ toward the states $|0\rangle$ and
$|\uparrow\downarrow\rangle$). On the other hand, there will be a transition
from the spin doublet to the singlet phase and therefore a mixing effect. Fig.
\ref{fig:correl_xi} shows the competition between the mixing effect (that
increases the correlations outside particle hole symmetry) and the evolution
toward the normal atomic limit (that lowers the correlations) if $\xi_d$ is
increased. The effect of the Coulomb interaction is once more found to favor the
single occupancy.

\subsection{\label{subsec:curr}Josephson current}
We now turn to the Josephson current through the quantum dot. The latter is
given by $J = 2e \frac{dF}{d\varphi}$ (where $F$ is the free energy). At zero
temperature, the free energy is the same than the level energies, so that the
Josephson current can readily be obtained once the renormalized energy levels
have been calculated.

Nevertheless, our analytical approach only describes the effective local limit
atomic states, and we can therefore only determine the current through the
Andreev bound states. Yet, it is known that the Josephson current also contains
a contribution of the continuum of states.\cite{Current_benjamin} The latter can
be of the same order and opposite sign as the bound state contribution.
Furthermore, Bauer at al.\cite{NRG_spectral_Bauer} have shown that the spectral
weight of the bound states may vary importantly as a function of the different
parameters (like the Coulomb interaction $U$), especially in the spin doublet
phase. As we exclusively investigate the effective local limit states, we do not
keep track of this effect as well. Therefore, the Josephson currents obtained in
our approach will only provide a rather rough and qualitative idea of the actual
total Josephson current.

\begin{figure}
\centering
\includegraphics[scale=0.25]{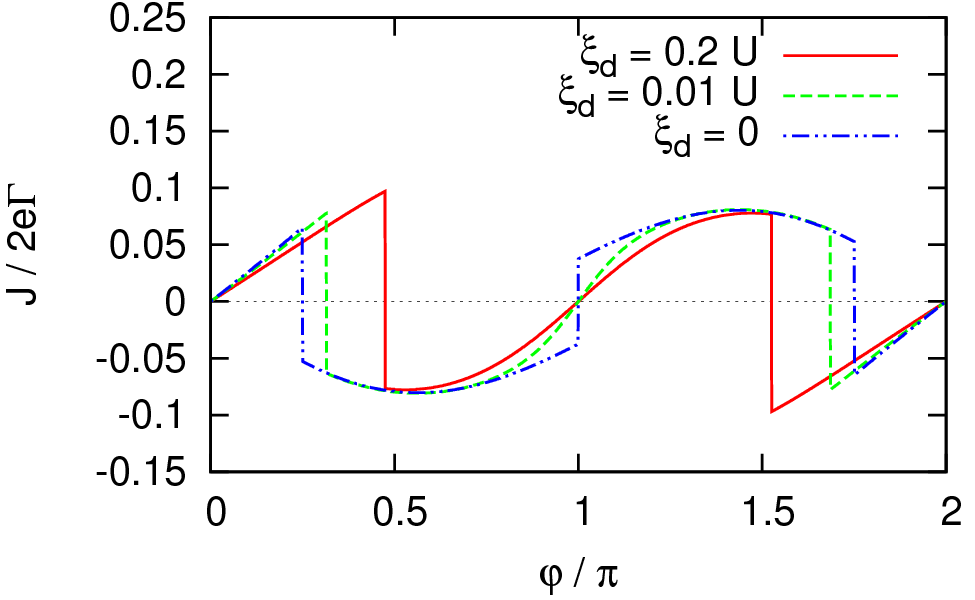}
\caption{(Color online) Josephson current through the bound states for $U = 3 \, \Gamma$, $\Delta = 0.1 \, D$.}
\label{fig:current_xi}
\end{figure}

Fig. \ref{fig:current_xi} shows the Josephson current calculated as the phase
derivative of the ground state energy $E_x$, $J = \frac{dE_x}{d\varphi}$, for
different values of $\xi_d$. One notices two regimes: If the phase is close to
$\varphi = 0$, the system will be in the BCS-like state. As there is no magnetic
moment in this phase, the ground state corresponds to a $0$-junction (i.e. phase
difference $\varphi = 0$). If $\varphi$ increases, the energy of the BCS-like
state increases (as can be understood in the effective local limit, where $E_- =
U/2 - \sqrt{{\xi_d}^2+{\Gamma_{\varphi}}^2}$). When the BCS-like state crosses
with the spin doublet, the ground state changes and the dot becomes singly
occupied. This magnetic moment leads to a discontinuous jump in the Josephson
current and the formation of a $\pi$-junction. Again, we notice that the spin
doublet is stabilized in the particle hole symmetric case. 

\section{\label{sec:conclu}Conclusion}

In this section we summarize our main results. First, it has been shown how
the Hamiltonian of a quantum dot coupled to superconducting leads can be mapped
onto an effective local model if the superconducting gap $\Delta$ is much bigger
than the characteristic energy scales of the dot. This limit can be quite
generally regarded as a low frequency expansion of the Green's function of the dot 
rather than the limit $\Delta \to \infty$ used in the literature.
This enabled us to extend the effective local Hamiltonian to leads with a finite 
electronic bandwidth.

We have then set up a perturbation theory around this local effective
Hamiltonian and established self-consistent equations for the energy
renormalizations of the Andreev bound states. We have derived those equations
based on the fact that the latter correspond to transitions between different
states of the local effective Hamiltonian.

In a last section, we used our formalism to calculate physical quantities such
as the Andreev bound state energies or superconducting correlations, and
understood how these evolve as a function of gate voltage, hybridization,
Coulomb interaction and superconducting gap amplitude. 
It has been shown that our simple approach agrees well with NRG
data in a vast range of parameters, with the main limitation that the Kondo 
temperature is not quantitatively described near particle-hole symmetry.
However, most experimentally interesting regimes should be described correctly
by the simple equations we have derived.

The simplicity and portability constitute the main advantages of our
approach, if one is interested in the Andreev bound states only, compared to 
extended numerical simulations. As the perturbative description is analytical and based 
on atomic-like levels, it should in principle be able to describe more complex 
systems like multiple quantum dots or molecules with several orbitals coupled to 
superconducting environments, and be readily applicable to describe future
spectrosopic measurements. Extensions of our formalism to the computation of the 
tunneling current at realistic gap values in three-terminal geometries \cite{Governale1} 
relevant for STM experiments should certainly deserve further scrutiny.

\begin{acknowledgments}
We wish to acknowledge stimulating discussions with D. Feinberg and C. Winkelmann, and
thank J. Bauer, A. Oguri and A. Hewson for providing us their NRG data.
\end{acknowledgments}

\appendix*
\section{\label{sec:delta_e_details}Derivation of the energy corrections}

The partition function is derived starting from the action's perturbation expansion in section \ref{sec:pert}. The actual calculations are performed in the operator formalism. It is very useful to note that the product of two fermionic (or bosonic) Greens functions $G_a(\tau)$ and $G_b(\tau)$ obeys $\int_0^{\beta}{d\tau\int_0^{\beta}{d\tau' G_a(\tau-\tau')G_b(\tau-\tau')}} = \beta \int_0^{\beta}{d\tau G_a(\tau)G_b(\tau)}$ (as can be shown using Fourier transformation). The partition function's perturbation expansion is

\begin{widetext}
\begin{eqnarray}
Z &=& Z_0 - Z_0 t^2\beta \sum_{\vec{k},i}{\int_{0}^{\beta}{d\tau \left(
G_{\vec{k}i\vec{k}i;11}^{0}(\tau)\langle T_\tau d_{\uparrow}^{\dagger}(\tau)d_{\uparrow}^{}(0)\rangle_0\right.}}\nonumber\\
&&\left.-G_{\vec{k}i\vec{k}i;12}^{0}(\tau)\langle T_\tau d_{\uparrow}^{\dagger}(\tau)d_{\downarrow}^{\dagger}(0)\rangle_0
- G_{\vec{k}i\vec{k}i;21}^{0}(\tau)\langle T_\tau d_{\downarrow}^{}(\tau)d_{\uparrow}^{}(0)\rangle_0
+ G_{\vec{k}i\vec{k}i;22}^{0}(\tau)\langle T_\tau d_{\downarrow}^{}(\tau)d_{\downarrow}^{\dagger}(0)\rangle_0\right)\nonumber\\
&&-2\beta\left|\Gamma_{\varphi}\right|\left( \langle T_\tau d_{\uparrow}^{\dagger}(0)d_{\uparrow}^{\dagger}(0)\rangle_0 + \langle T_\tau d_{\downarrow}^{}(\tau)d_{\uparrow}^{}(0)\rangle_0 \right) \text{ .}
\end{eqnarray}
\end{widetext}
In the above equation, $G_{\vec{k}i\vec{k}i;ij}^{0}$ is the Fourier transformed Nambu matrix element $\left.\widehat{G}_{\vec{k}i,\vec{k}i}^{0}(i\omega_n)\right|_{i,j}$ and the subscript $0$ indicates that the expectation values are statistical averages calculated in the effective local limit. The leads' Green's functions are:

\begin{eqnarray*}
\sum_{\vec{k}}{G_{\vec{k}i\vec{k}i;11}^{0}(\tau)} &=& 
\sum_{\vec{k}}{-\frac{\operatorname{sgn}(\tau)}{2}\left(e^{-|\tau| E_{\vec{k}}}+e^{-(\beta-|\tau|)E_{\vec{k}}}\right)}\\
G_{\vec{k}i\vec{k}i;12}^{0}(\tau) e^{-i\varphi_i} &=& \frac{\Delta}{2E_{\vec{k}}}\left(e^{-|\tau| E_{\vec{k}}}-2\cosh(|\tau| E)n_F(E_{\vec{k}})\right)\\
&\stackrel{T \to 0\,\mathrm{K}}{\to}& \frac{\Delta}{2E_{\vec{k}}}\left(e^{-|\tau| E_{\vec{k}}}-e^{-(\beta-|\tau|)E_{\vec{k}}}\right)
\end{eqnarray*}
with $E_{\vec{k}} = \sqrt{{\epsilon_{\vec{k}}}^2+\Delta^2}$. Furthermore, $G_{\vec{k}i\vec{k}i;21}^{0}(\tau) = {G_{\vec{k}i\vec{k}i;12}^{0}(\tau)}^*$ and $\sum_{\vec{k}}{G_{\vec{k}i\vec{k}i;22}^{0}(\tau)} = \sum_{\vec{k}}{G_{\vec{k}i\vec{k}i;11}^{0}(\tau)}$.

As one cannot apply Wick's theorem because of the Coulomb interaction, the dot's Green's functions are calculated using Lehmann representation, which yields (for $\tau > 0$)
\begin{widetext}
\begin{eqnarray}
\langle T_\tau d_{\uparrow}^{\dagger}(\tau)d_{\uparrow}^{}(0)\rangle_0 &=& \frac{1}{Z_0} \left(u^2\left(e^{-E_-^0\tau}e^{-E_{\uparrow}^0(\beta-\tau)}+e^{-E_{\downarrow}^0\tau}e^{-E_+^0(\beta-\tau)}\right)\right.\nonumber \\
&&\left.+v^2\left(e^{-E_+^0\tau}e^{-E_{\uparrow}^0(\beta-\tau)}+e^{-E_{\downarrow}^0\tau}e^{-E_-^0(\beta-\tau)}\right)\right)\\
\langle T_\tau d_{\uparrow}^{\dagger}(\tau)d_{\downarrow}^{\dagger}(0)\rangle_0 &=& \frac{1}{Z_0}uv \left(e^{-E_{\downarrow}^0\tau}e^{-E_-^0(\beta-\tau)}-e^{-E_-^0\tau}e^{-E_{\uparrow}^0(\beta-\tau)}\right.\nonumber \\
&&\left.-e^{-E_{\downarrow}^0\tau}e^{-E_+^0(\beta-\tau)}+e^{-E_+^0\tau}e^{-E_{\uparrow}^0(\beta-\tau)}\right)\\
\langle T_\tau d_{\downarrow}^{}(\tau)d_{\uparrow}^{}(0)\rangle_0 &=& \frac{1}{Z_0}uv \left(e^{-E_{\downarrow}^0\tau}e^{-E_-^0(\beta-\tau)}-e^{-E_-^0\tau}e^{-E_{\uparrow}^0(\beta-\tau)}\right.\nonumber \\
&&\left.-e^{-E_{\downarrow}^0\tau}e^{-E_+^0(\beta-\tau)}+e^{-E_+^0\tau}e^{-E_{\uparrow}^0(\beta-\tau)}\right)\\
\langle T_\tau d_{\downarrow}^{}(\tau)d_{\downarrow}^{\dagger}(0)\rangle_0 &=& \frac{1}{Z_0} \left(u^2\left(e^{-E_+^0\tau}e^{-E_{\uparrow}^0(\beta-\tau)}+e^{-E_{\downarrow}^0\tau}e^{-E_-^0(\beta-\tau)}\right)\right.\nonumber \\
&&\left.+v^2\left(e^{-E_-^0\tau}e^{-E_{\uparrow}^0(\beta-\tau)}+e^{-E_{\downarrow}^0\tau}e^{-E_+^0(\beta-\tau)}\right)\right) \text{ .}
\end{eqnarray}
\end{widetext}

Using $u^2+v^2=1$, the partition function becomes:
\begin{widetext}
\begin{eqnarray}
Z &=& Z_0 + \beta t^2 \sum_{\vec{k},\sigma} \nonumber\\
&&\times\left(\frac{1}{E-(E_+^0-E_{\sigma}^0)}\left(e^{-\beta E_+^0}-e^{-\beta (E+E_{\sigma}^0)}\right) + \frac{1}{E-(E_-^0-E_{\sigma}^0)}\left(e^{-\beta E_-^0}-e^{-\beta (E+E_{\sigma}^0)}\right)\right.\nonumber\\
&&+ \frac{1}{E+(E_+^0-E_{\sigma}^0)}\left(e^{-\beta E_{\sigma}^0}-e^{-\beta (E+E_+^0)}\right) + \frac{1}{E+(E_-^0-E_{\sigma}^0)}\left(e^{-\beta E_{\sigma}^0}-e^{-\beta (E+E_-^0)}\right)\nonumber\\
&&+ \frac{2\Delta}{E}uv\left|\cos(\frac{\varphi}{2})\right| \nonumber \\
&&\times\left(\frac{1}{E+(E_+^0-E_{\sigma}^0)}\left(e^{-\beta E_{\sigma}^0}-e^{-\beta(E+E_+^0)}\right)-\frac{1}{E+(E_-^0-E_{\sigma}^0)}\left(e^{-\beta E_{\sigma}^0}-e^{-\beta(E+E_-^0)}\right) \right.\nonumber\\
&&\left.\left.- \frac{1}{E-(E_+^0-E_{\sigma}^0)}\left(e^{-\beta E_+^0}-e^{-\beta (E+E_{\sigma}^0)}\right) + \frac{1}{E-(E_-^0-E_{\sigma}^0)}\left(e^{- \beta E_-^0}-e^{-\beta (E+E_{\sigma}^0)}\right)\right)\right) \nonumber \\
\label{Z_delta_inf_tot}
&&+2\beta\left|\Gamma_{\varphi}\right|uv\left(e^{-\beta E_+^0}-e^{-\beta E_-^0}\right)
\end{eqnarray}
\end{widetext}

As $E_{\vec{k}} = \sqrt{{\epsilon_{\vec{k}}}^2+\Delta^2} > 0$, terms with an $e^{-\beta E_{\vec{k}}}$ are exponentially suppressed for $T \to 0 \, \mathrm{K}$ and can be omitted.

\end{document}